**Title:** Study and calculation of thermal conductance of thermal infrared detectors using finite element method

**Authors:** P. Ramos, A. Andrés; A. López and A. Manzanares

**Address:** GRIFO. Departamento de Electrónica, Universidad de Alcalá, Alcalá de Henares, E-28871 Madrid, Spain. **Email:** pablo.ramos@uah.es

**Abstract:** In this paper, a multilayer thermal infrared detector model has been achieved by finite element method (FEM). All contributions of the thermal conductance were taken into account and calculated. In order to maximize the detector response, it is necessary to reduce the thermal conductance. Dynamic simulation in 3D was used to optimize this FEM model. The effect of the substrate properties of the detector on its response has been studied. Moreover, different boundary conditions have been analyzed. Optimal detector response values are obtained when the substrate thermal conductivity and its thickness are small. Moreover, a vacuum packing of the detector will be necessary to increase the detector responsivity.

**Highlights**

A multilayer thermal infrared detector model has been achieved by FEM.

Dynamic simulation in 3D was used to optimize the model.

G decreases as the substrate thermal conductivity and its thickness increase

A vacuum packing of the detector will be necessary to increase the detector response.

**Introduction**

The solid state infrared thermal detectors are built with materials that have a physical property (electrical resistance, electric charge, etc), which is a function of temperature. The absorption of infrared radiation produces a temperature change in the detector. Its technological applications include the detection of infrared radiation for thermal imaging, monitoring of medical procedures, test thermal dissipation in integrated circuits, etc [1, 2].

In the optimization of a thermal detector, the thermally sensitive material is an important factor but it is not the unique. Geometric factors (the detection area, the thickness of material ...) and thermal factors (thermal losses [3], the packages in a chip ...) are essentials for the development of a good detector.

The infrared thermal detectors are described by the classical bolometer equation [4]

$$H(d\Delta T/dt) + G\Delta T = F\eta P_i(t) \tag{1}$$

In this expression it is assumed that $F\eta P_i(t)$ is the power absorbed during a lapse of time, $dt$, quickly delivered in the whole volume of the material (area $A$ and thickness $d$), being $H$, $G$, $F$ and $\eta$ the thermal capacity, thermal conductance, IR-filter transmittance and emissivity of the detector surface, respectively. $P_i(t)$ is the modulated radiation that is desired to be measured.

The voltage responsivity ($R_V$) is the figure of merit that determines the electrical response of the detector due to the incident radiation.

$$R_V = \frac{\Delta V}{\Delta P_i} \quad (V/W) \tag{2}$$

where $\Delta V$ is the output voltage and $\Delta P_i$ is the incident radiation.

The voltage responsivity of the detector, which is obtained by solving the bolometric equation, is given by the following expression:

$$R_v = \frac{K\eta}{G\sqrt{1+\omega^2\tau^2}} \qquad (3)$$

Where $K = \Delta V / \Delta T$ is a coefficient that reflects the ability of the material to transduction of changes in temperature ($\Delta T$) in an output voltage of detector [1].

The thermal time constant ($\tau$) is defined by the following expression:

$$\tau = \frac{H}{G} \qquad (4)$$

In order to increase the responsivity of a detector, it is necessary to reduce thermal losses of sensitive material, because $R_v$ is inversely proportional to the thermal conductance ($G$). Actually, to reduce the thermal losses, the substrate is micro-mechanised in floating membrane, closed membrane or cantilever form, by means of a controlled chemical etching. However, the speed response of detector decrease, because $\tau$ grows (equation 4). In fact, it is necessary a good adjustment of the thermal conductance value to optimize both the voltage responsivity and the thermal time constant.

The thermal conductance can be calculated by solving complicated differential equations of heat transfer, or with electrical equivalent circuits which determine the thermal conductivity of detector [5].

L.E. Great *et al* [3], using a thermal losses model for infrared detectors, achieved the voltage responsivity which is inversely proportional to the thermal conductance.

Numerical techniques used to solve differential equations such as finite element method (FEM). This technique allows you to evaluate the three-dimensional spatial variations of thermal losses and find out the parts of the detector with the greatest losses. Also, it is possible to perform simulations in various physical domains [6].

There are some references in this sense, in which a FEM model is used to calculate the static thermal conductance of bolometers thermal detectors [7,8]. It is obtained as the ratio between heat flux and temperature variation on the surface between sensible material and the substrate. However, these models are not described in detail, and they do not give the dynamic response and vertical conductance of the detector either [9]. Moreover, these studies are only simulated the lateral conductance which is important to avoid cross-talk in thermal images. The finite element method has proved to be a useful tool for detector design, providing a very good fits between simulated and experimental data of conductance [10]. There are few works that use finite element method to design of pyroelectric detectors [11, 12].

.

The aim of this paper is to calculate the total conductance of the bolometer thermal detectors and not just lateral conductance by finite element method. We present here a dynamic FEM model in three-dimensional. The developed model is a thermal multilayer model with closed membrane substrate. For different values of the model parameters, the variation of substrate thermal conductance (*G*) values, which are obtained by simulation, was achieved. Actually, the values of thermal conductivity, the thickness of the substrate and the boundary conditions were modified. Finally, we discuss and compare the simulation results with experimental data of a pyroelectric thermal detectors.

**Experimental method and Fem model**

Figure 1 shows a cross section view of the structure of standard infrared thermal detector. A sensitive material, which was deposited onto substrates, absorbs infrared radiation on the circular top electrode. In order to achieve an improvement in the performance of the detector, the bottom of the substrate etches to reduce its thickness and heat losses of the detector through the substrate (membrane closed).

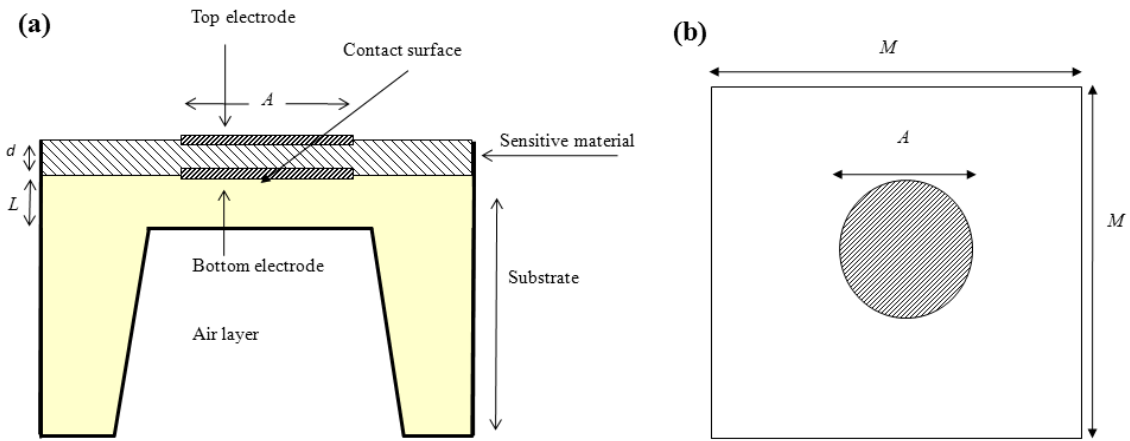

**Figure 1.** Schematic diagram of a standard thermal infrared detector. a) Cross view of the detector's structure and b) its top view.

The FEM model consists of three parts (figure 2): the substrate, sensitive material and surrounding air (top and bottom). The substrate has a thickness $L$ and at its bottom was applied two different boundary conditions. The first was air conduction (layer thickness of 1mm) and the second was air convection ($h = 5$ W m$^{-2}$ K$^{-1}$ and $T_{Air}=$ 300 K). Initially, the whole detector was at 300 K. At the top of the substrate, the boundary condition was air thermal conduction or vacuum. In the contact surface between the substrate and sensitive material is where heat losses occur (figure 1). In this contact surface was applied a temperature square wave, from 300 K to 310 K at a frequency of oscillation of 0.25 Hz. Finally, this square wave was applied at the sensitive element for study the heat losses effect with sensitive element and without it.

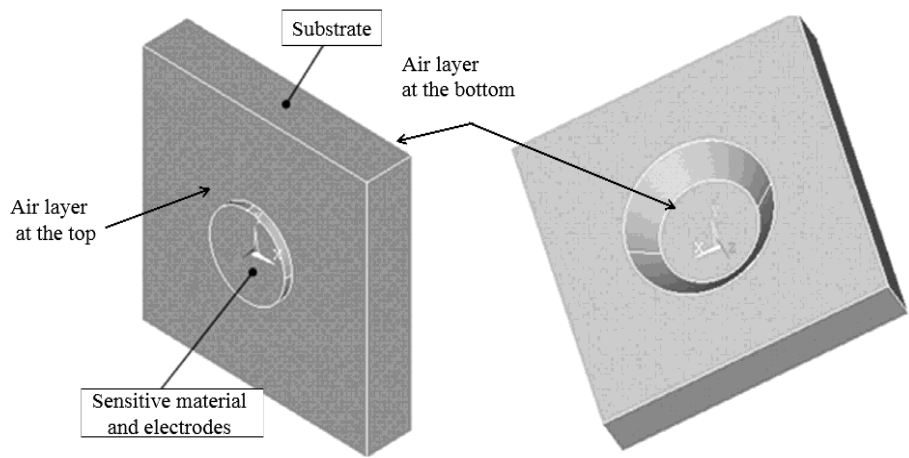

**Figure 2**. Basic structure of the tri-layer FEM model.

The conduction heat transfer equation has been solved by the finite element method from the circular surface of contact to the bottom air layer through the substrate. The above boundary conditions at the top and the bottom of detector have been applied. The material properties of the sensitive element, the substrate and the air have been included in the FEM model (density, specific heat and thermal conductivity). A mesh with pyramidal elements has been used because it fits better to the circular geometry (figure

3).

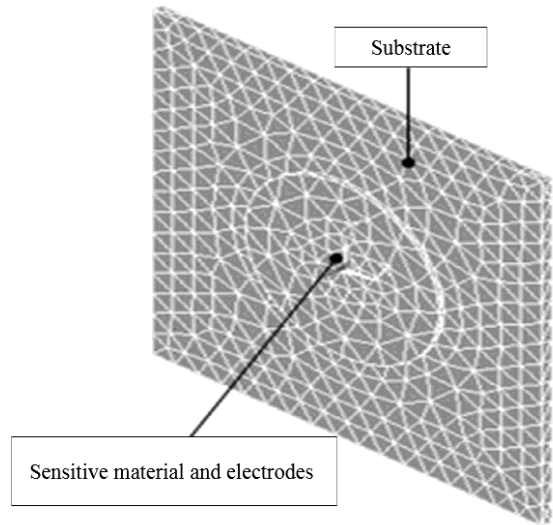

**Figure 3**. Mesh of the tri-layer FEM model of the detector

The simulation results are the temperature spatial distribution of the whole detector (figure 4). An analysis of these spatial distribution on the substrate, it be obtained that the bottom of the substrate has temperatures closer to the initial stimulus (310 K), deducing that the greatest heat losses are produced in the substrate. There are also some thermal losses at the top of the substrate. Both losses have been analyzed by measuring the heat flux through the contact surface between the sensitive material and the substrate.

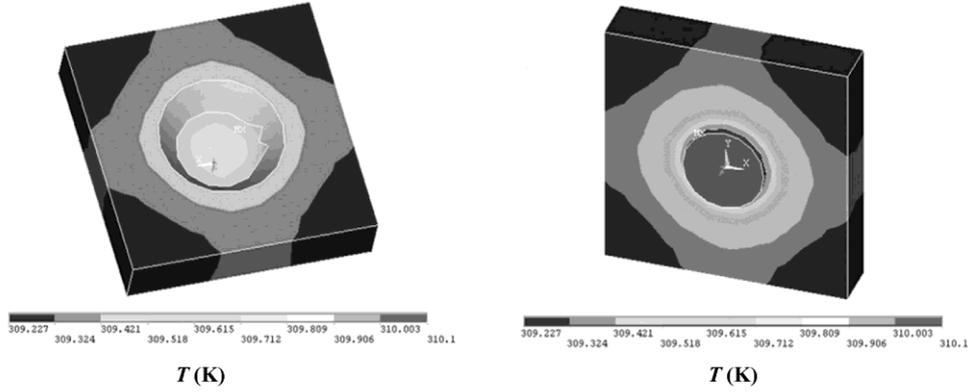

**Figure 4.** Temperature spatial distribution on the tri-layer FEM model.

In figure 5 a dynamic study of the total heat flux for different values of thermal conductivity of the substrate is shown. It must be noted that the total $\dot{Q}_{tot}$ heat flux, increases to a maximum and then it decreases exponentially. The equation 5 fits this behavior.

$$\dot{Q}_{tot}(t) = \dot{Q}_0 + \dot{Q}_1 e^{-(t-t_0)/\tau} \tag{5}$$

where $\dot{Q}_0$ and $t_0$ are adjustment constants.

Finally, values for $\tau$ have been calculated by adjustment of simulated heat flow with equation 5. The thermal conductance $G$ of the substrate has been obtained by dividing the heat capacity $H$ of the substrate by the calculated $\tau$.

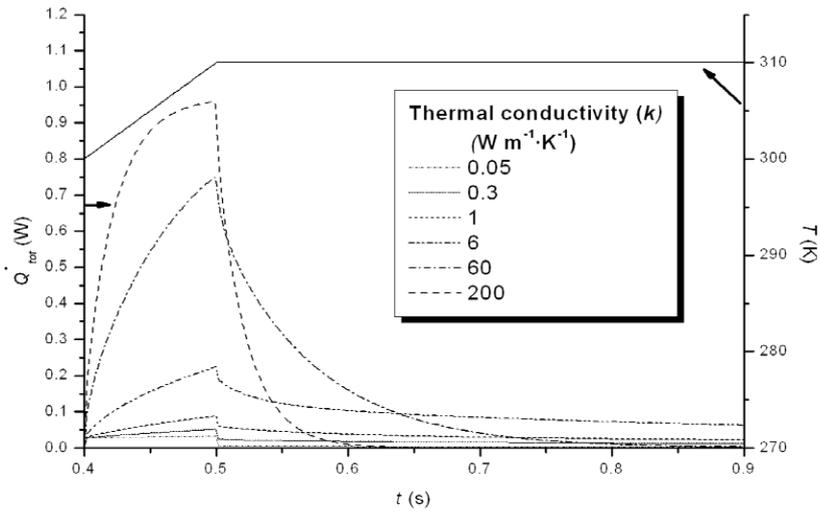

**Figure 5.** Total heat flux vs time, varying the substrate thermal conductivity for bi-layer FEM model. Boundary conditions at the bottom and top of the substrate are air convection and vacuum, respectively

**Results and discussion**

Bi-layer FEM model has been carried out. This model consists of two layers: a layer of substrate with thickness $L$, and a bottom layer of air (1mm). The temperature boundary conditions were 310 K on the circular surfaces of the substrate, where the sensitive material must be deposited, and 300 K on the bottom of the air layer. The heat conduction equation has been solved by FEM. The static thermal conductance, $G$, has been calculated using the ratio of simulated heat flux through the substrate and its simulated temperature difference. Table I shows the simulation results of the bi-layer model for different thicknesses of substrate, being noticeable that increasing substrate thickness causes an increase in thermal conductance. This effect may be attributed to the different values of thermal conductivities. The substrate has more thermal conductivity than the air, thus generating more thermal losses (thermal sink effect).

**Tabla I.** Bi-layer FEM model. Simulated static thermal conductance, $G$ against substrate thickness, $L$ with a substrate thermal conductivity of 60 Wm$^{-1}$K$^{-1}$.

| $L$(μm) | $G$(W/K) |
|---|---|
| 200 | 2.92 10$^{-4}$ |
| 300 | **3.34 10$^{-4}$** |
| 400 | 3.89 10$^{-4}$ |
| 500 | 4.67 10$^{-4}$ |

In figure 6 the variation of thermal time constant values, which were obtained by dynamic simulation, for different values of the substrate thickness is shown as a function of temperature oscillation frequency ($f$). As it can be seen, the values of the thermal time constants decrease as frequency is increased. Actually, the simulated value of $\tau$ is equal to 68 ms for a substrate value of 300 μm and a frequency of 0.25 Hz that is very close to the static frequency.

Using the static simulated value of $G$ (table I) and the dynamic value of $\tau$ (figure 6), the heat capacity $H$ was calculated by equation 4, being equal to 2.27 10$^{-5}$ $J/K$ for a substrate thermal conductivity of 60 Wm$^{-1}$K$^{-1}$.

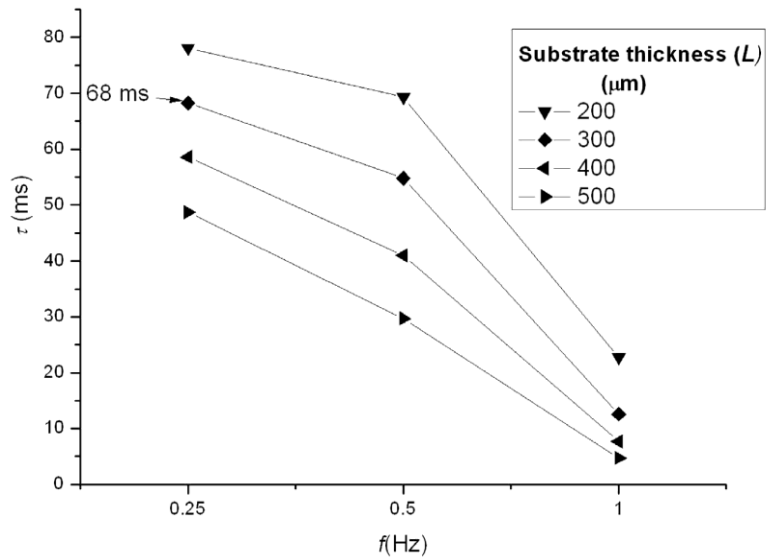

**Figure 6.** Thermal time constant against the frequency for bi-layer FEM model. Boundary conditions at the bottom and top of the substrate are air convection and vacuum, respectively.

In figure 7 the variation of $\tau$ values obtained by dynamic simulation for a substrate thickness of 300 µm is shown as a function of substrate thermal conductivity ($k$). The value of $\tau$ was calculated according to equation 5 for two different boundary conditions that were applied at the bottom of the substrate (air convection and air conduction).

It must be noted that $\tau$ decreases as $k$ increases, because $G$ is a quasi-proportional function to $k$ (see equation 4). The results for both boundary conditions are very similar. Actually, the values of simulated $\tau$ converge to $k$ greater than one, which is the value for most common substrates.

The effects of air conduction on top of the detector have been simulated with two boundary conditions: air conduction and vacuum (figure 8). As heat loss increases through the top of the detector, $\tau$ values are lower for simulation with a layer of air. Moreover, the difference in values of thermal time constant is 5% for both boundary conditions.

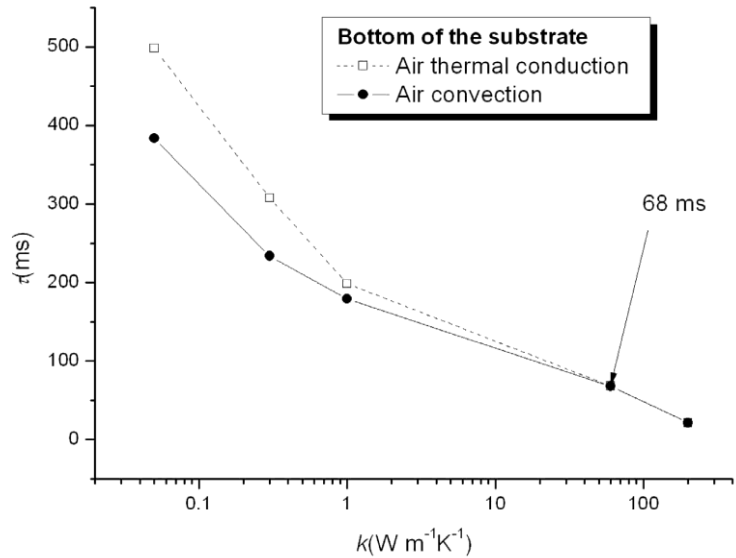

**Figure 7.** Thermal time constant vs thermal conductivity for bi-layer FEM model. Boundary conditions at the bottom and top of the substrate are two different conditions (air thermal conduction vs convection) and vacuum, respectively.

The previous bi-layer model has been used to simulate the thermal time constant with different boundary conditions that were applied at sensitive material-substrate interface. These simulations only take into account the behavior of the substrate. A tri-layer model was achieved, introducing an extra layer that corresponds to the sensitive material and electrodes (figure 2). In this case, the boundary conditions were applied on top of the new layer. The effect of sensitive material on the $\tau$ response for different values of substrate thickness are shown in figure 9, where low values for time thermal constant can be seen (approximately 10%). This difference is due to the lower values of the thermal capacity of the bi-layers model for simulations with tri-layers model (equation 6)

$$H^{-1}_{Tri-layers} = H^{-1}_{Bi-layers} + H^{-1}_{Extra-layer} = \frac{1}{2.1*10^{-5}} \left(K/J\right) \quad (6)$$

with [13] $H_{Extra-layer}$=2.25 $10^{-4}$ J/K y $H_{Bi-layers}$=2.27 $10^{-5}$ J/K

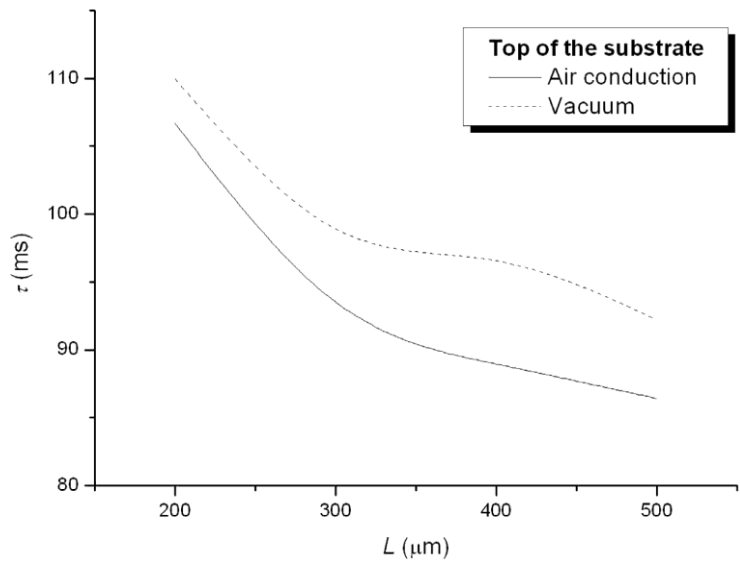

**Figure 8.** Thermal time constant against substrate thickness for bi-layer FEM model. Boundary conditions at the bottom of the substrate is air thermal conduction and at top of the substrate are air conduction and vacuum, respectively.

In order to verify the similarity between experimental data of infrared detector and simulated response of tri-layer FEM model, the thermal conductance ($G_{FEM}$) is represented in figure 10, being noticeable that increasing the substrate thickness causes an increase in *G*. Simple thermal conduction model (TCM) and data of a pyroelectric detector are used to calculate the values of thermal conductance. The pyroelectric detector [14] is a thin film of nominal composition $(Pb_{0.76}Ca_{0.24})TiO_2$ that was deposited by spin coating onto an etched substrate of MgO(100) (similar to figure 1). Top and bottom platinum electrodes were deposited by sputtering using a mask of circular electrodes, 1 mm of diameter, arranged in a matrix of two columns and three rows.

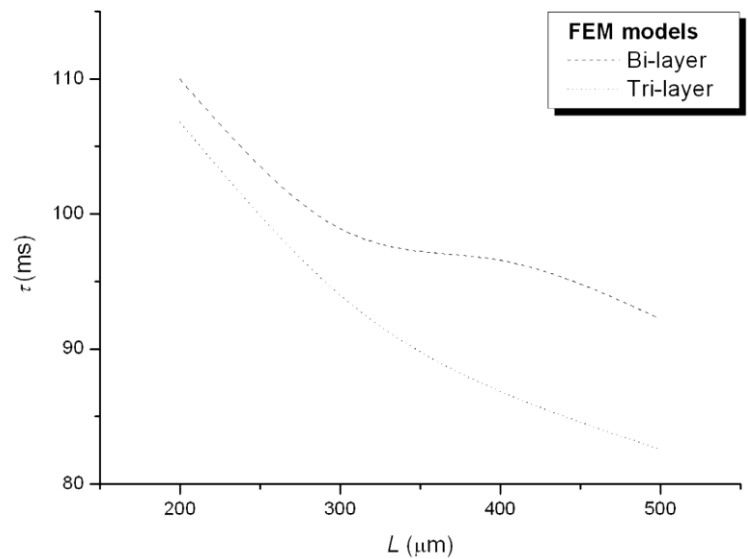

**Figure 9** Thermal time constant vs substrate thickness for bi-layer and tri-layer FEM model. Boundary conditions at the bottom and top of the substrate are air convection and vacuum, respectively

Using the steady-state solution of one-dimensional heat conduction equation [15] for a bi-layer structure (substrate and air layers), thermal conductance of this simple thermal conduction model ($G_{TCM}$) is calculated by equation 7. The thickness of substrate (*L*) and the air layer ($L_{air}$) are variable. However, the total thickness of the bi-layer structure is constant (1mm).

$$\frac{1}{G_{TCM}} = \frac{1}{G_{sub}} + \frac{1}{G_{air}} \qquad (7)$$

where

$$G_{air} = \frac{k_{air} S}{L_{air}} \qquad G_{sub} = \frac{k_{sub} S}{L} \qquad S = \pi \left(\frac{A}{2}\right)^2 \qquad (8)$$

$k_{air}$ and $k_{sub}$ is the thermal conductivity of air and the substrate, respectively.

$A$ is diameter of the electrode.

Operating, the value of thermal conductance is equal to:

$$G_{TCM}(W/K) = \frac{S k_{sub} k_{air}}{L_{air} k_{sub} + L k_{air}} = \frac{S k_{sub} k_{air}}{(10^{-3} - L) k_{sub} + L k_{air}} = \frac{S k_{sub} k_{air}}{k_{sub} 10^{-3} - L |k_{sub} - k_{air}|} \qquad (9)$$

An approximation of the thermal conductance is calculated by equation 9. Only the vertical contribution of the thermal conductance is considered. The developed tri-layer FEM model could evaluate both vertical and lateral contributions.

The experimental thermal conductance ($G_{exp}$) were adjusted to the experimental results of pyroelectric detectors with different thicknesses of substrate, using a Pspice model of the thermo-electric response [16].

The thermal conductance of the tri-layer FEM model ($G_{FEM}$) is calculated by the thermal time constants (figure 9), using equation 4 and the value of $H_{FEM}$ (equation 6).

As it can be seen in figure 10, the $G_{FEM}$ of finite element simulation and $G_{exp}$ of the pyroelectric detector and the $G_{TCM}$ of simple model were plotted against the substrate thickness.

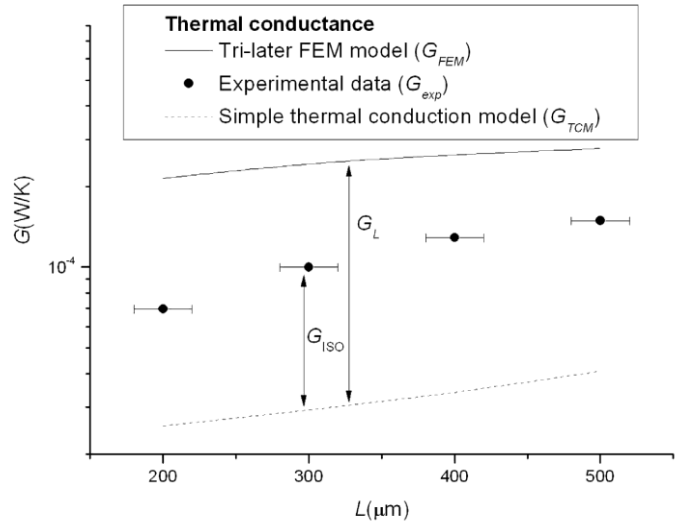

**Figure 10.** Thermal conductance against substrate thickness for different models. The models are tri-layer FEM model (FEM), simple thermal conduction model (TCM) and experimental data of pyroelectric detector (EXP).

$G_{FEM}$ is larger than $G_{TCM}$, because simple thermal model only took into account the vertical thermal losses, while the tri-layer FEM model calculated both vertical and lateral losses (equation 10). The experimental thermal conductance values are between the above models. The FEM model only has been considered one sensitive material and the experimental detector was an array of sensitive elements that were partially isolated from each other. In equations 12 and 13, this isolation effect is modeled by the lateral thermal conductance without isolation ($G_L$) in parallel with a lateral thermal conductance with isolation effect.

$$G_{FEM} = G_V + G_L \qquad (10)$$

$$G_{TCM} = G_V \tag{11}$$

$$G_{exp} = G_V + G_{LISO} \tag{12}$$

$$G_{LISO}^{-1} = G_L^{-1} + G_{ISO}^{-1} \tag{13}$$

where

$G_V$ and $G_L$ are the vertical and lateral thermal conductance, respectively.

$G_{LISO}$ and $G_{ISO}$ are total lateral thermal conductance with thermal isolation and only thermal conductance with isolation effect, respectively.

In this case, it is concluded that from the above equations calculating it is possible to calculate each contribution of total thermal conductance (see table II), which is a determinant factor when optimizing the detector response.

**Tabla II.** Tri-layer FEM model. The different contribution of the total thermal conductance, $G$ against substrate thickness, $L$ with a substrate thermal conductivity of 60 Wm$^{-1}$K$^{-1}$.

| $L$(μm) | $G_V$(W/K) | $G_L$(W/K) | $G_{ISO}$(W/K) |
|---|---|---|---|
| 200 | 2.55 10$^{-5}$ | 1.90 10$^{-4}$ | 5.78 10$^{-5}$ |
| 300 | 2.92 10$^{-5}$ | 2.16 10$^{-4}$ | 1.05 10$^{-4}$ |
| 400 | 3.40 10$^{-5}$ | 2.30 10$^{-4}$ | 1.62 10$^{-4}$ |
| 500 | 4.08 10$^{-5}$ | 2.36 10$^{-4}$ | 2.00 10$^{-4}$ |

It is found that vertical thermal losses are lower than lateral one that will be reduce to decrease the total thermal conductance, *G*. Thus, the detector voltage responsivity must be made as large as possible if a good performance is desired (equation 3).

**Conclusions**

The tri-layer FEM model allows dynamic and spatial analysis of the thermal detectors to easily change its design parameters such as geometry, physical and isolation properties. Moreover, finite element simulations of thermal detectors predict reliably their temporal response.

Optimal voltage responsivity values are obtained when the total thermal conductance is small. In order to reduce the total thermal conductance, it is necessary to use a substrate with small thermal conductivity and thickness, and a package that allows the substrate to have a layer without air (vacuum) on its top and under its bottom. Furthermore, the lateral isolation is essential to maximize the response of the thermal detector.